\documentclass[manuscript]{aastex6}

\usepackage{graphicx}
\usepackage{natbib}
\usepackage[english]{babel}  
\usepackage{subfigure}
\usepackage{url}
\newcommand{\new}{ }
\newcommand{\neww}{ }
\newcommand{\newb}{ }
\newcommand{\newc}{ }

\begin{document}

\title{X-ray flare oscillations track plasma sloshing along star-disk magnetic tubes in Orion star-forming region}



\author{Fabio Reale\altaffilmark{1}}
\affiliation{Dipartimento di Fisica \& Chimica, Universit\`a di Palermo,
              Piazza del Parlamento 1, 90134 Palermo, Italy;
              fabio.reale@unipa.it}
\author{Javier Lopez-Santiago}
\affiliation{Department of Signal Theory \& Communications, Universidad Carlos III de Madrid, Avda. de la Universidad 30, Leganes, 28911 Madrid, Spain}

\author{Ettore Flaccomio, Antonino Petralia, Salvatore Sciortino}
\affiliation{INAF-Osservatorio Astronomico di Palermo, Piazza del Parlamento 1, 90134 Palermo, Italy}

\altaffiltext{1}{and INAF-Osservatorio Astronomico di Palermo, Piazza del Parlamento 1, 90134 Palermo, Italy}

\begin{abstract}
Pulsing X-ray emission tracks the plasma echo traveling in an extremely long magnetic tube that flares in an Orion Pre-Main Sequence (PMS) star.  On the Sun, flares last from minutes to a few hours and the longest-lasting typically involve arcades of closed magnetic tubes. Long-lasting X-ray flares are observed in PMS stars. Large-amplitude ($\sim 20$\%) long-period ($\sim 3$~hours) pulsations are detected in the light curve of day-long flares observed by the {\it Advanced CCD Imaging Spectrometer} (ACIS) on-board {\it Chandra} from PMS stars in the Orion cluster. Detailed hydrodynamic modeling of two flares observed on V772~Ori and OW~Ori shows that these pulsations \newb{may} track the sloshing of plasma along a single long magnetic tube, triggered by a sufficiently short ($\sim 1$~hour) heat pulse. This magnetic tubes are as long ($\geq 20$~solar radii) as to connect the star with the surrounding disk. 
\end{abstract}


\section{Introduction} 
\label{sec:intro}

Close to the end of their formation, stars are surrounded by a gas and dust disk, from which planets form. Magnetic fields are known to play a key role in the star-disk system \citep{Johns-Krull2014a}. It is believed that the inner regions of the disk are significantly ionized by the stellar radiation and that accreting material flows along magnetic channels that connect the disk to the star \citep{Koenigl1991a}. 
Very long and intense X-ray flares in star forming regions might occur in such long channels \citep{Favata2005a}, but this is highly debated \citep{Getman2008a}. 
These flux tubes might resemble those observed in the solar corona and diagnosed in the stellar coronae, but on a much larger scale. On the Sun we see the so-called coronal loops on the scale of several thousand kilometers in active regions, but some faint large-scale structures can extend up to $\sim 1$~$R_\odot$ \citep{Reale2014a}. Most solar flares occur in active region loops, but the long-lasting ones can involve more and more loops aligned in arcades. The other stars are so distant that we cannot resolve the flaring regions, but it is supposed that they occur in loops and even in arcades. Whereas the duration of solar flares typically ranges from few minutes to several hours, stellar flares can be very intense, more than the solar bolometric luminosity, and long-lasting, more than one day, in very active stars. Several of such gigantic coronal flares have been surveyed in star-forming regions \citep{Favata2005a} and where they occur is a big question. Magnetic instabilities in flux tubes were proposed as the origin of the flaring activity also in T~Tauri stars \citep{Birk1998a,Birk2000a}, and long-lasting stellar flares might be expected to involve loop arcades \citep{Getman2008a} as on the Sun. 
In the long-lasting solar flares, the duration is mainly due to the progressive involvement of more and similar loops, and therefore it is not directly linked to the size of the flaring structures. This might be the case also for the giant stellar flares. 
However, if a single stellar loop were flaring, the cooling time of the confined plasma would be proportional to the loop length \citep{Serio1991a,Reale2014a}, and day-long flares would correspond to giant loops, as long as possibly connecting the star with the disk \citep{Hartmann2016a}. 
There are ways to distinguish between a pure cooling in a single loop and a decay only due to progressive reduction of the energy release in a loop arcade \citep{Reale1997a}, but the arguments are debated and the uncertainties are large \citep{Getman2008a}. 
Several studies \citep{Favata2005a,Giardino2007a} find results compatible with long magnetic channels in PMS stars, but the derivation of the loop length is based on the assumption of a flare occurring in a single loop \citep{Reale2007b}. 

Different diagnostics independent of flare cooling would be desirable to solve this ambiguity between long and arcade flaring structures. A new way would be to detect and study brightness wave fronts traveling along the magnetic flux tubes. These might prompt periodic pulsations in the flare light curves, whose period would track back to the length of the wave guide. This concept has been pursued recently in a general hydrodynamic modeling framework and used as a new tool to diagnose the duration of the flare heat pulse \citep{Reale2016a}. Periodic pulsations have been detected in the light curves of several stellar X-ray flares \citep{Mitra-Kraev2005a,Welsh2006a,Pandey2009a,Lopez-Santiago2016a}, but generally not linked to the size of the flaring structure.

\neww{In this work, we show that the modulated light curve of long flares observed in the soft X-rays with Chandra/ACIS on young stars (V772 Ori, OW Ori) in the Orion star-forming region is well explained by flaring plasma sloshing back and forth inside a single flaring magnetic tube, that is several stellar radii long. Such long structures can be hardly mapped as semicircular loops, and in the following we will address them more generally as magnetic flux tubes.
Both stars are surrounded by a disk and have been observed to be active accretors (see Section~\ref{sec:data} for more details).} The length of the flaring tubes analysed in this work is enough to suggest a connection between the star and the accretion disk. 

Section~\ref{sec:data} describes the data,  in Section~\ref{sec:model} they are modelled and the results discussed in Section~\ref{sec:discus}.

\section{The data analysis}
\label{sec:data}

For this work, we use data from the Chandra Orion Ultradeep Project \citep[COUP,][]{Getman2005a}. {We focus on V772~Ori (COUP~43) and on OW~Ori (COUP~1608), two young M stars that underwent strong, long-duration flares during the observations \citep{Favata2005a}.}

{V772 Ori \citep[$A_\mathrm{V} = 1.18$,][]{Aarnio2010a} is an M1.5 star \citep{Hillenbrand2013a}  
with M=0.4\,M$_\odot$ and radius R$_\star$=2.9\,R$_\odot$ \citep{Getman2005b}. 
It was listed as a candidate double-lined spectroscopic binary \citep{Rhode2001a}, still not confirmed \citep{Biazzo2009a}.
For this star a photometric period $P = 1.69 \pm 0.02$~days was measured and it was classified as an accretor based on its intense H$_\alpha$ emission line \citep{Stassun1999a}. Equivalent widths were determined for $H_\alpha$ and Li~I from high-resolution ($R > 30000$) optical spectra \citep{Stassun1999a}. The $EW(\mathrm{H}\alpha)$ was corrected for possible narrow nebular emission line. The results were: $EW(\mathrm{Li~I}) = 260$~m\AA, $EW(\mathrm{H}\alpha) = 43.4$~\AA \citep[see Fig.7 in][]{Stassun1999a}. Radio VLA fluxes (F(4.5GHz) $= 0.88 \pm 0.27$~mJy and F(7.5GHz) $= 0.67 \pm 0.18$~mJy) also confirm the presence of a disk around this star \citep{Kounkel2014a}. }

 
{The total COUP light curve shows two flares for this star, one after the other \citep{Favata2005a}. We study the second one, which is observed in its entirety, while for the first one we only see part of the decay. There is no particular evidence that one flare is related to the other, so we will assume they are not.} 

{OW~Ori is a well-known M0.5e classical T Tauri star located inside the Orion 
Nebula Cluster \citep{Hillenbrand1997a}, with mass 
M=0.48\,M$_\odot$ and radius R$_\star$=1.77\,R$_\odot$ \citep{Getman2005a}.
Infrared excesses $\Delta(I-K) = 1.41$ \citep{Hillenbrand1998a} and $\Delta(H-K) = 0.13$ \citep{Getman2008a} were derived.
Both are indicative of an accretion disk. The extinction towards OW~Ori 
is $A_\mathrm{V} = 0.25$\,mag \citep{Aarnio2010a}.}
\neww{It was derived \citep{Aarnio2010a} a dust destruction radius in the accretion disk 
$R_\mathrm{dust} = 10$~R$_\odot$, close to the co-rotation radius ($R_\mathrm{cor} = 
8.6$~R$_\odot$) \citep{Getman2008a}. These results indicate the
presence of an inner disk. The rotational period of the star is 
$P = 4.21$~days \citep{Rhode2001a}.}

The X-ray characteristics of both flares were studied previously \citep{Favata2005a, Getman2008a}. We focus our attention on the oscillation patterns in the X-ray light curve revealed by the \emph{Chandra X-ray Observatory} (CXO) during the stellar flares.
{Data reduction was performed within the COUP. Details can be found in \citet{Getman2005a}.} The Chandra dither has \newc{negligible} effect on the sources analysed in this work. During the analysis performed for \cite{Lopez-Santiago2016a}, we found the $\sim 700$~s pattern due to Chandra’s dither in some stars of the COUP close to a gap between chips. However, this is not the case for COUP 43 or COUP 1608 \citep{Getman2005a}, \newc{whose large off-axis angle makes the counts distribution large compared to the area of the dead columns}. 


Spectral analysis for the flare of V772~Ori {was} carried out independently in two works \citep{Favata2005a,Getman2008a}, using different methods.  \cite{Favata2005a}  {performed} a time-resolved spectral analysis {by} extracting X-ray spectra in time intervals selected through a maximum likelihood algorithm, with the pre-imposed condition of having enough photon statistics for spectral analysis. 
{Spectral fits of the resulting X-ray spectra were then done using the XSPEC  package assuming the MEKAL spectral emissivity model for coronal equilibrium plasma.} The  obtained plasma characteristics in each time interval are mean values of the actual plasma parameters during that time period. 
%
The spectral range analyzed in this work {was} $0.5-7$~keV. A single thermal component {was} assumed, with the global abundance fixed to $0.3~Z_\odot$ \citep{Favata2005a}. The absorbing column density {was} left free to vary. With these constraints, a temperature of 58~MK {was} found for the flaring plasma, corresponding to a peak temperature $T_{peak} = 142$~MK \citep{Favata2005a,Reale2007a}. The peak emission measure is $EM_{peak} \approx 10^{54}$~cm$^{-3}$ and the absorption column density is $\log N_H \mathrm{[cm^{-2}]} \approx 21$.

Instead,  \cite{Getman2008a} {based their} analysis on the method of adaptively smoothed median energy (MASME). In practice, they {analyzed} the median energy of photons during the flare and the count-rate instead of performing a time-resolved spectral fitting. Despite the two works obtain different results for the peak temperature and emission measure for some stars of the COUP, this is not the case for our target (V772~Ori).  \cite{Getman2008a} {determined} $EM_{peak} = 1.2 \times 10^{54}$~cm$^{-3}$ and $T_{obs} = 59$~MK, which are very close to the values {derived} by  \cite{Favata2005a}. 

The raw light curve is actually dominated by low count statistics ($\sim 0.03-0.04$ cts/s) and the presence of any periodic signal is hardly detectable from direct inspection and with no data interpolation. 
To reveal oscillations in the light curve of V772~Ori in an objective way, we follow the procedure applied to other COUP flares \citep{Lopez-Santiago2016a}. Firstly, we select events for the X-ray source only in the time period of the flare and generate a light curve with temporal resolution {100 seconds}. Then, this flare light curve is normalized using a moving average. With this procedure, the general shape of the flare is subtracted. The normalized flare light curve is then convolved with a Morlet function, which has the advantage of preserving information in the time domain \citep{Lopez-Santiago2016a}. The result of the convolution is a 2D power spectrum: the time-frequency representation of the light curve. If any oscillation is present in the original dataset, it is revealed as a feature (peak) in the wavelet power spectrum. The significance of such peaks is determined by generating confidence levels. To do this, a background noise model must be assumed. {This background noise is derived from the univariate lag-1 autoregressive, or Markov process. Details
on the methodology are given in \cite{Torrence1998a} and summarized in \cite{Lopez-Santiago2016a}}. In our work, we assume \emph{red noise} instead of the white noise background typically used in the literature. The assumption of white noise for deriving confidence levels in a flare light curve usually overestimates the significance of low-frequency patterns and underestimates the significance of high-frequency patterns. 
{\newb{The time binning of 100 s is not an integral multiple of the ACIS frame time for these observations, i.e., 3.14104~s, but this produces no significant aliasing in our results. We also repeated our analysis with other sampling periods, i.e., 50~s, 30~s, 5~s, and obtained the same results. }

\begin{figure}
   \centering
   \includegraphics[width=10cm]{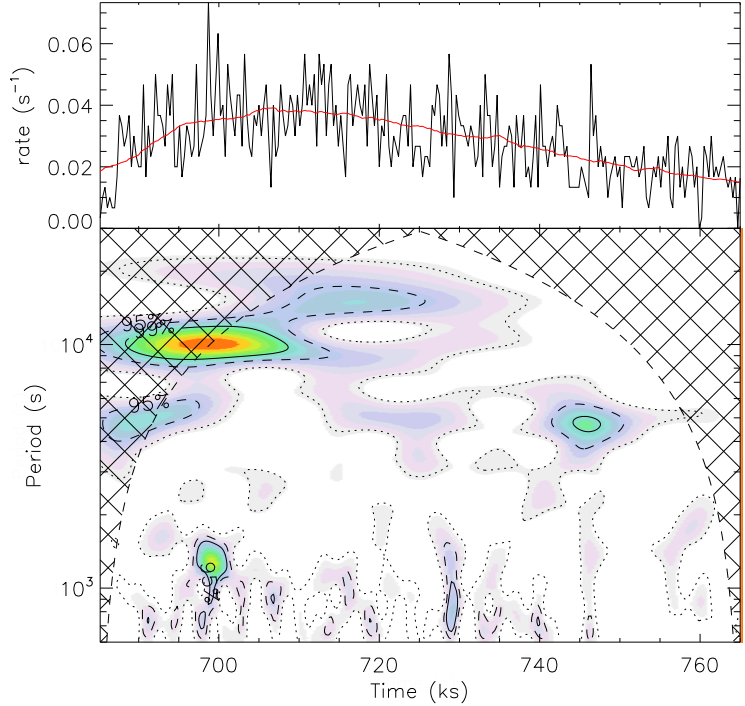} 
   \caption{Flare light curve (upper panel) and wavelet power spectrum (bottom panel)
   for V772~Ori. The continuos line in the upper panel represents the moving average 
   used to subtract the general shape of the flare. The contours in the bottom panel 
   are confidence levels: {continuous line is 99\%, dashed line is 90\% and 
   dotted line is 67\%.} The hatched area is the cone of influence (COI), the region 
   of the wavelet power spectrum in which edge effects become important.}
   \label{fig:wavelet}
\end{figure}
%


\newc{This work addresses the detection of pulsations with the longest duration, which may be evidence of very long loops. We consider as actual (long) pulsations only those features lasting for at least three times the period at which they are revealed in the power spectrum.}
Figure~\ref{fig:wavelet} presents the result of the wavelet transform for the flare underwent by the star V772~Ori during the COUP observations. The upper panel shows the light curve analyzed in this work. The line overplotted is the moving average (bin = 15~ks). The bottom panel shows the wavelet power spectrum. Contours are for the confidence levels 67\% (dotted line), 95\% (dashed line) and 99\% (continuous line), assuming red noise as the background signal. The power spectrum reveals a significative \newb{persistent} feature at period $P = 10 \pm 1$~ks ($\sim 3$~hrs) \newb{over a time range of $\sim 40$~ks} around the peak of the flare. This feature corresponds to a pulsation extending for approximately four periods, as actually shown in Figure~\ref{fig:lc}.


\begin{figure}              
 \centering
 \subfigure[]
   {\includegraphics[width=11cm]{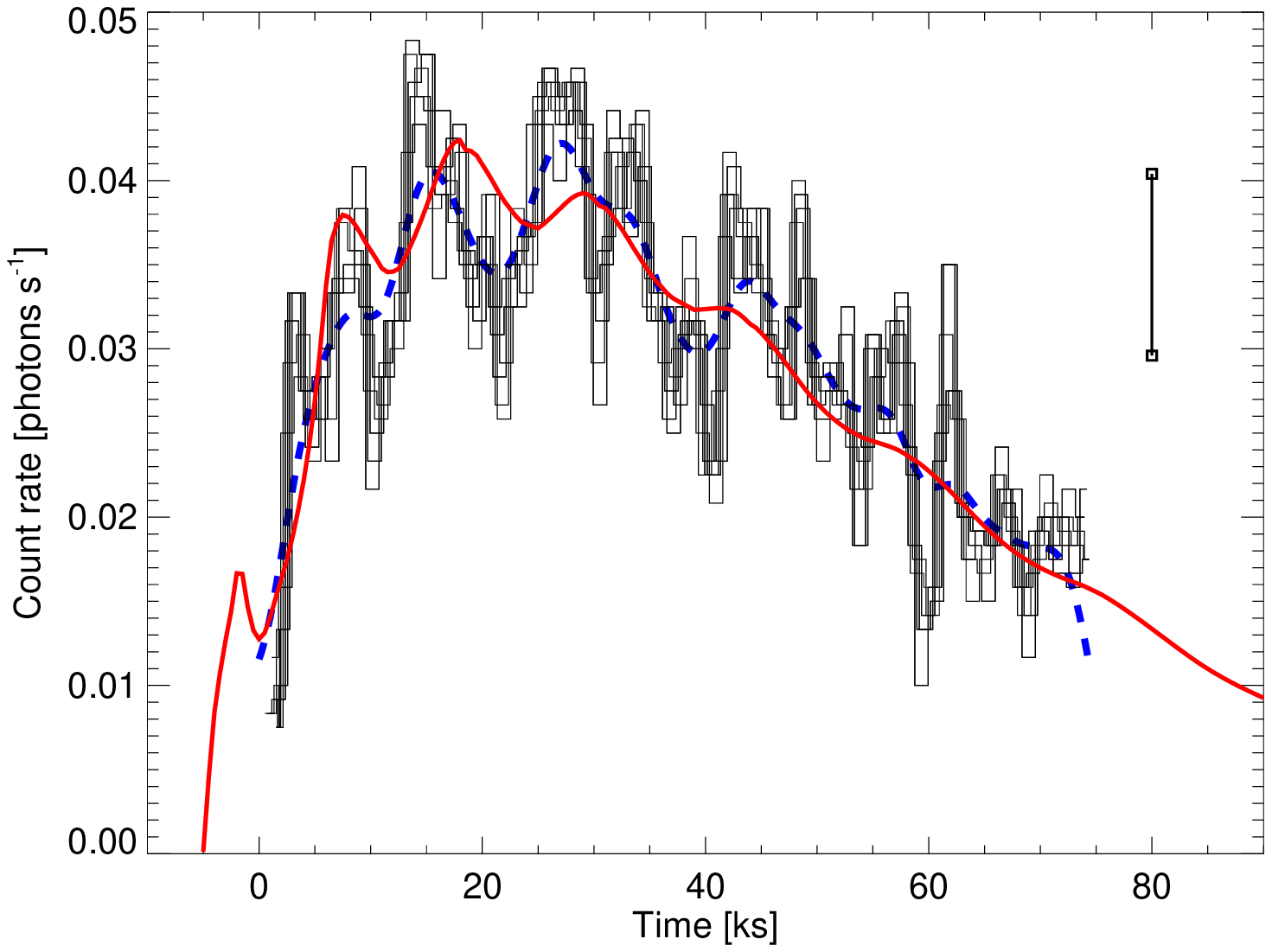}}
 \subfigure[]
   {\includegraphics[width=11cm]{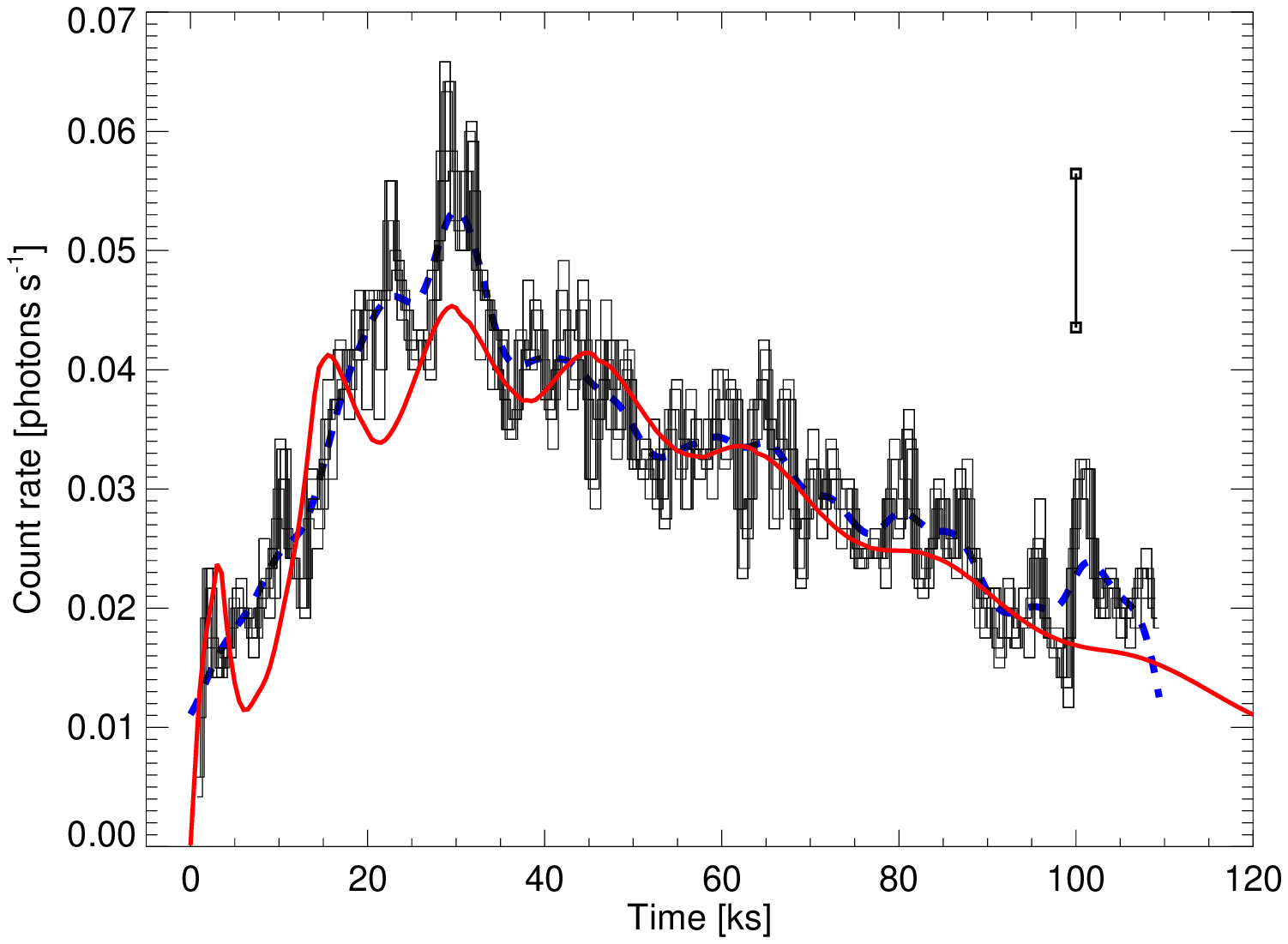}}
\caption{\footnotesize \neww{Flare light curves observed with Chandra/ACIS from (a) V772~Ori {(COUP~43; Obs. ID 4374)} and (b) OW~Ori {(COUP~1608; Obs. ID 4373)}, {with a binning of 1200~s, each shifted by 100~s} {(black solid)} and after smoothing with a Gaussian with $\sigma = 2000$~s {(blue dashed)}. The observed light curves are compared to those obtained from respective hydrodynamic simulations of a flaring flux tube 20 and 30 $R_\odot$ long and heat pulses of $\sim 1$ hour {(red solid)}. } The vertical bar on the right marks a typical data error bar.} 
\label{fig:lc}
\end{figure}

{Figure~\ref{fig:lc}a shows the observed light curve {with a binning of 1200~s, replicated after slight shifts of 100~s} (black solid line),  and after smoothing with a Gaussian with $\sigma = 2000$~s (blue dashed line). {Each binned light curve is noisy but the envelope of them \newb{provides a hint of} the dips and cusps of the oscillations. The smoothed light curve shows periodic pulsations with the period found with the wavelet analysis above. We point out that the detection of the oscillations is made from the light curve with the fine binning shown in Fig.~\ref{fig:wavelet}. 

{A possible interaction between binary components certainly cannot explain \newb{the period of} the observed oscillations: according to Kepler's laws, the period is too short to be compatible with the orbital period of a possible binary, unless the system is very unlikely a contact binary.}

{For the flare on OW~Ori, the wavelet analysis \citep{Lopez-Santiago2016a} revealed a feature at $P \sim 10$~ks. The authors showed that this feature corresponds to 
an oscillation with amplitude $\geq 5$\% of the flare intensity ($\Delta I / I \geq 0.05$). Figure~\ref{fig:lc}b shows the related light curve, in the same format as for the flare of V772~Ori. }

\section{Hydrodynamic modeling}
\label{sec:model}

\neww{As in previous stellar flare modeling \citep{Reale1988a,Favata2005a,Testa2007a,Schmitt2008a}, we model the flaring plasma as confined inside a magnetic flux tube.}  The plasma is compressible and confined by the magnetic field and moves and transports energy along the field lines. The plasma evolution can then be described with a single-fluid one-dimensional hydrodynamic model where the coordinate is the distance along the tube. We assume that the tube is a closed coronal flux tube anchored \new{at the footpoints}. In a typical corona the footpoints are anchored at two different locations of the chromosphere/photosphere. In the case of a young star one of (or even both) the locations might be on the disk which might be {as dense as a stellar chromosphere} \citep{Kastner2002a,Telleschi2007a,Argiroffi2007a}. The model is analogous to that used in  \cite{Favata2005a}. The magnetic tube is assumed symmetric with respect to the middle, and we model only half of it. The gravity component along the flux tube is computed assuming a radius $R_* = 3 R_\odot$ and a surface gravity $g_*= 0.1 g_\odot$, typical of low-mass PMS objects \citep{Favata2005a}. \neww{A fine tuning of these parameters is not important for the modeling because the plasma evolution is largely dominated by the heating and cooling processes and by the dynamics driven by the heat pulse \citep{Bradshaw2010a}. In the following, we describe in detail the model for the flare on V772~Ori.}

The length of the tube is mainly determined by the flare decay time which -- for a closed flaring coronal magnetic flux tube starting from equilibrium conditions -- scales as  \citep{Serio1991a,Reale2014a}:

\begin{equation}
\tau_d \approx 5 \frac{L_\odot}{\sqrt{T_6}}
\label{eq:taud}
\end{equation}
where $\tau_d$ is in hours, $T_6$ is the flare maximum temperature in units of $10^6$ K, $L_\odot$ is the tube total length in units of solar radii ($R_{\odot}$). 
Since the thermal conduction is extremely efficient along the tube, we can estimate its maximum temperature from the scaling laws for static coronal loops \citep{Rosner1978a,Reale2014a}, even if the tube is never close to an equilibrium during the flare. We obtain:

\begin{equation}
T_6 \sim 2.3 H_{-3}^{2/7} L_\odot^{5/21} \sim 180 
\label{eq:tmax}
\end{equation}
where $H_{-3}$ is the heating rate per unit volume in units of $10^{-3}$~erg~cm$^{-3}$~s$^{-1}$.
\newb{Although the observed decay time is about 4 hours (Figure~\ref{fig:lc}a) which, for this estimated temperature, corresponds to a length $L_\odot \sim 12$, after test simulations, we find that a tube length $L_\odot = 20$ matches better the observed light curve, i.e., the decay and the oscillation period.}

We assume that there is plasma confined in this long tube before the flare. This plasma is much cooler and more tenuous than it gets during the flare, with a maximum temperature of 17~MK and a pressure of 2~dyn~cm$^{-2}$ at the footpoints, similar to previous work \citep{Favata2005a}. These initial conditions are kept steady and at equilibrium in the corona by a uniform heating of $3.5 \times 10^{-5}$~erg~cm$^{-3}$~s$^{-1}$. The tube atmosphere includes a relatively thick chromosphere at the footpoints which is described according to one standard model \citep{Vernazza1981a}. \new{In typical flare loop modeling the chromosphere has mainly the role of providing a mass reservoir for filling the flux tube with dense plasma \citep{Reale2014a}. So we assume that this model chromosphere is valid  both for footpoints rooted on the stellar surface and on the disk.}
The flare is triggered in the tube atmosphere by suddenly releasing a powerful heat pulse, much stronger than the equilibrium heating mentioned above.  The condition to trigger the plasma sloshing inside the magnetic tube, which determines the  pulsations observed in the light curve, is that the duration of the heat pulse $\tau_H$ must be shorter than return sound crossing time along the tube $\tau_s$ at the peak of the flare \citep{Reale2016a}:

\begin{equation}
\tau_H < \tau_s \approx 1.5 \frac{L_\odot}{\sqrt{T_6}}
\label{eq:sct}
\end{equation}
where $\tau_H$ and $\tau_s$ are in hours.
The pulsation period is still the sound crossing time but taken during the flare decay, i.e., longer than $\tau_s$, because in Equation~(\ref{eq:sct}) we should use quite a smaller temperature than the flare maximum temperature $T_6$. On the other hand, since these flares, as many others observed in young stars, are very hot, reaching temperatures above 100~MK \citep{Favata2005a}, according to Equation~(\ref{eq:sct}), a measured period of $\sim 3$~hours involves tube lengths $\geq 20 R_{\odot}$. Moreover, since the decay time in Equation~(\ref{eq:taud}) scales as the sound crossing time in Equation~(\ref{eq:sct}), we will always see a similar number of pulsations in this and other analogous flares, i.e., about five-to-ten \citep{Reale2016a}.

%

With the maximum temperature $T_6$ derived in Eq.(\ref{eq:tmax}), from Eq.(\ref{eq:sct}) we obtain $\tau_H < 2.5$~hrs. We obtain a very similar evolution with heat pulses with a duration of 3000-4000~s ($\approx 1$~hour), safely within the condition to trigger the plasma sloshing, and deposited in the corona either uniformly along the tube with an intensity of 0.1~erg~cm$^{-3}$~s$^{-1}$, i.e. $\sim 3000$ times stronger that the equilibrium heating, or more concentrated toward the middle of the tube (a Gaussian with a width $\sigma_H = 10^{11}$ cm) and an intensity 0.5 erg cm$^{-3}$ s$^{-1}$. The heat distribution is not an important parameter because the thermal conduction is very efficient and levels out the temperature in a short time along the tube \citep{Bradshaw2006a}.

\begin{figure}               
 \centering
   {\includegraphics[width=14cm]{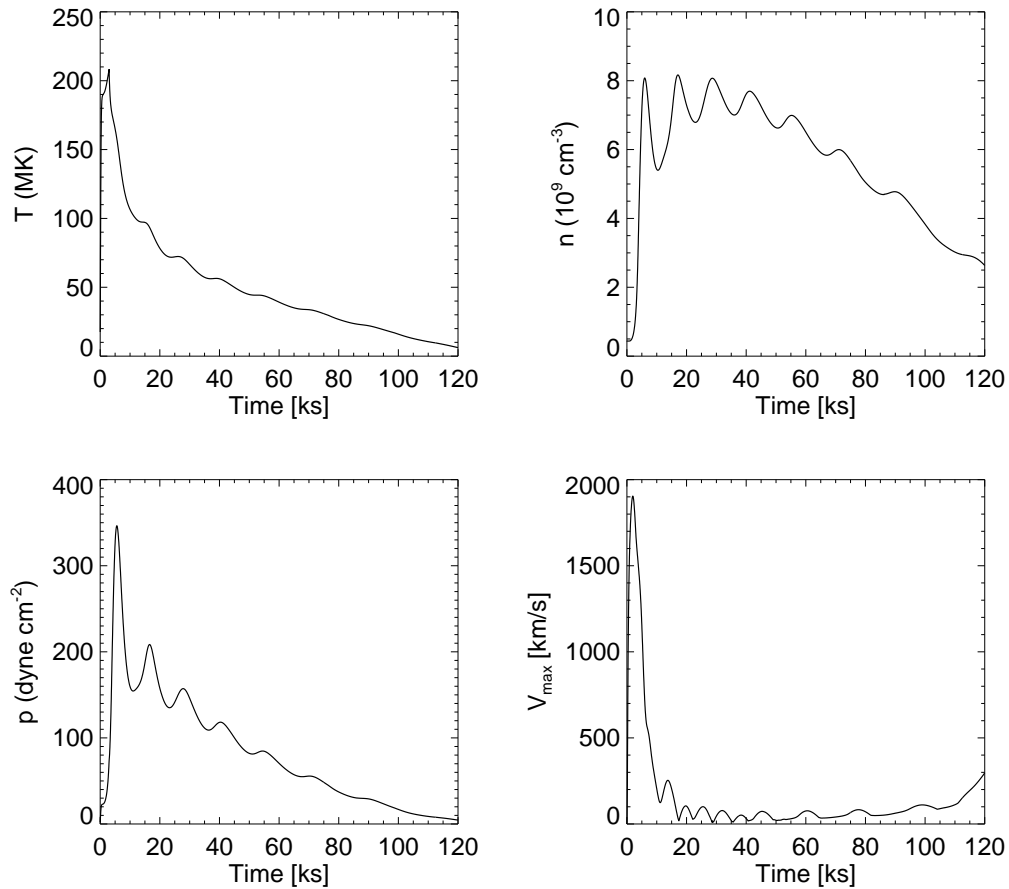}}
\caption{\footnotesize Evolution of the temperature $T$, density $n$, pressure $p$ at the middle of the flaring tube, and of the maximum plasma speed $V$ along the tube for the simulation of the flare on V772~Ori. }
\label{fig:top}
\end{figure}

We let the plasma confined in the tube evolve from the initial conditions above and under the action of the heat pulse above, by solving numerically the hydrodynamic equations for a compressible plasma. The equations and numerical model are described in previous works \citep{Peres1982a,Betta1997a}. The grid is adaptive with a maximum spatial resolution of 0.1~km. 
Figure~\ref{fig:top} shows the evolution of some representative quantities for 120~ks, i.e., the temperature, density and pressure at the middle of the tube, and the maximum absolute speed reached by the plasma along the tube. The temperature rises fast to over 200~MK while the heat pulse is on, and then decreases rapidly and monotonically with an e-folding time $\tau_d \sim 4$~hours, not far from the conduction cooling time \citep{Reale2014a}. The pressure and density have a very different evolution. In particular, the density rises more slowly, up to $n \sim 10^{10}$~cm$^{-3}$, then it remains relatively steady for about 50 ks and finally decreases gradually.


\begin{figure}              
\centering
  {\includegraphics[width=12cm]{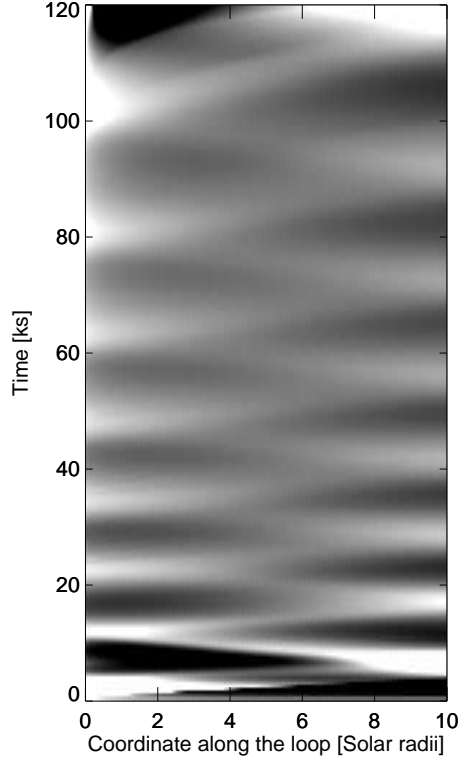}}
\caption{\footnotesize Evolution of the pressure along half of the flux tube from the flare simulation for V772~Ori. The pressure $\pm 20$\% of the mean at each time is shown. \neww{The gray scale is linear (black is low, white is high)}.}
\label{fig:pimg}
\end{figure}

This overall trend is modulated by well-defined pulsations, similar to those described in a previous work \citep{Reale2016a} and due to plasma sloshing back and forth along the magnetic tube. The period of the pulsations is determined by the sound crossing time during the cooling. In this case, a typical temperature in the decay is $\sim 50$~MK, which corresponds to $\tau'_s \sim 4$~hours. We see that the period increases with time as expected, because the temperature decreases continuously during the decay. The pressure at mid-point shows a well-defined peak $p_{th} \approx 350$ dyne cm$^{-2}$ at $t \approx 1.5$~hr, and then decreases with periodic pulsations. The velocity shows an initial very high peak of $v_{max} \sim 2000$~km/s due to the explosive expansion of the dense chromospheric plasma upwards into the much more tenuous corona. After this transient, also the velocity oscillates, as the density and pressure. The density and velocity determine the presence of a ram pressure hitting the tube footpoints on the order of $p_{ram} \approx n m_H v^2$, where $m_H$ is the mass of the hydrogen atom. For $n \sim 10^{10}$~cm$^{-3}$ and $v \sim 1000$ km/s, we obtain $p_{ram} \sim 200$ dyne~cm$^{-2}$, i.e., of the same order as the thermal pressure $p_{th}$ estimated above.

As described in  \cite{Reale2016a}, the plasma sloshing is driven by a depression which forms low in the tube, where the plasma cools more efficiently, as soon as the heat pulse stops. Figure~\ref{fig:pimg} shows the sloshing for this specific situation.

From the density and temperature distribution of the plasma along the model flux tube, we can synthesize the expected plasma X-ray spectrum filtered through the ACIS-I spectral response. We have computed the emission in the coronal part of the flux tube, i.e., above the transition region, as that of an optically thin plasma:

\begin{equation}
I(t) = A \int_L n(t)^2 G[T(t)] dL  ~~ {\rm cts ~ s^{-1}}
\label{eq:intens}
\end{equation}
where $t$ is the time, $n$ is the plasma density, $G(T)$ is the instrumental response function, i.e., the expected {count rate} spectrum at a temperature $T$ folded through the ACIS-I effective area, integrated over the energy range, {and scaled for the star distance}, $A$ is the tube cross-section. We assume a hydrogen column density $\log N_H \mathrm{[cm^{-2}]} = 21.2$ and less than solar (0.3) metal abundances \citep{Favata2005a}. The photon counts are integrated above 0.4 keV. The tube cross-section is a free parameter that is obtained from scaling the model light curve to the observed one. A good matching is obtained by assuming a circular cross section with a radius $R \approx 0.06$~L, not far from  values typical of solar flaring loops \citep[$R/L \sim 0.1$,][]{Golub1980a}. This corresponds to a cross-section area of $\sim 10$\% of the solar surface ( $\sim 1$\% of the actual surface of this star according to known parameters). In this case, for a such a long flux tube, the cross-section might change significantly along the tube and the value we obtain is to be intended as an average. A non-uniform cross-section does not lead to a significant change of the results obtained for a 1D description with uniform cross-section. \new{With this assumption, we obtain an emission measure $\approx 10^{54}$ cm$^{-3}$ at the flare peak, in very good agreement with the one diagnosed from the observation \citep{Favata2005a}. The maximum temperature in the range 100-150 MK in the flare rise is also in good agreement with the measured effective maximum temperature between 50 and 100 MK.} 

From Eq.~(\ref{eq:intens}) we derive the light curve shown in Fig.~\ref{fig:lc} (red solid line); it is not noisy because it is not affected by photon statistics (its sampling time is 500~s, much less than the binnings used for the data in the figure). As a check for consistency, we add Poisson noise to the model count rate and apply the same wavelet analysis as to the real data. 
\newc{Figure~\ref{fig:pois}c shows the light curve with the same binning as the observed one and the related $\Delta C$ statistics with respect to the fitted polynomial baseline, to be compared to an observed one (Fig.~\ref{fig:pois}a) among those shown in Fig.~\ref{fig:lc}a. }
The result of the wavelet analysis with a binning of 100~s is shown in Fig.~\ref{fig:pois}d: we find a strong feature at the long period $P \approx 10$~ks, and in a time range $\sim 40$~ks, around the peak of the flare, very similar to that detected in the data (Fig.~\ref{fig:wavelet}). \newc{The feature at $\sim 30$~ks is of short duration, i.e., less than 3 periods, and therefore not relevant in this work.} The wave reconstruction \citep{Torrence1998a,Lopez-Santiago2016a} in  Fig.~\ref{fig:pois}b shows data and model oscillations with the same period and amplitude of the same order (within $\sim 30\%$). 
\newc{Note that a non-sinusoidal but periodic signal would produce a wavelet power spectrum very different from that of a sinusoid. For example, the power spectrum of a periodic train of short pulses is revealed in the power spectrum as another pulse train in frequencies, i.e., a vertical feature repeated in time at the position of each pulse. This is not observed in our data. Our data present a wavelet power spectrum typical of damped sinusoids \citep[see Figure 9 of][for an example of an impulse train-like signal]{Addison2016a}. }

\begin{figure}              
 \centering
 \subfigure[]
   {\includegraphics[width=6cm]{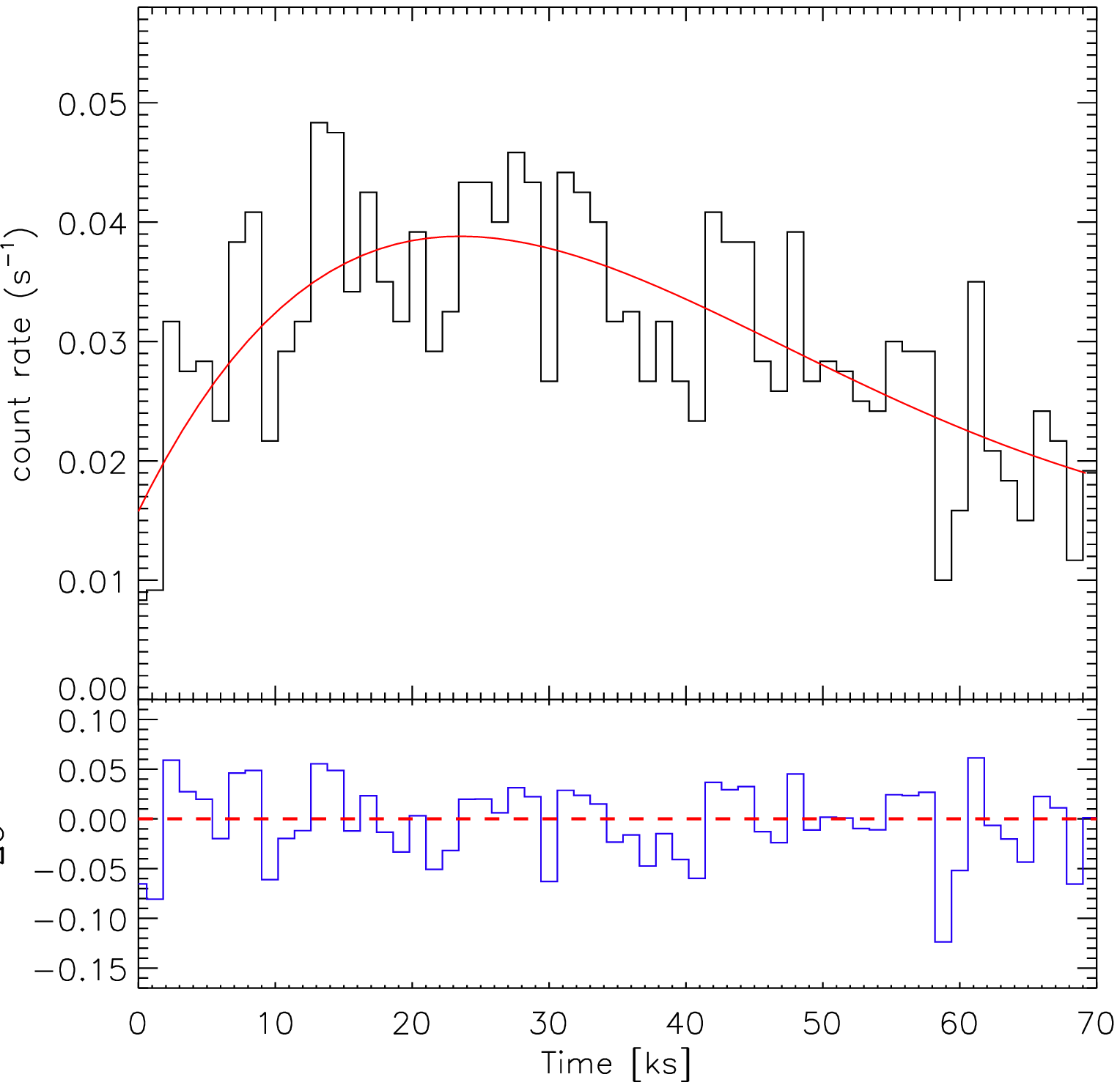}}
\subfigure[]
   {\includegraphics[width=6cm]{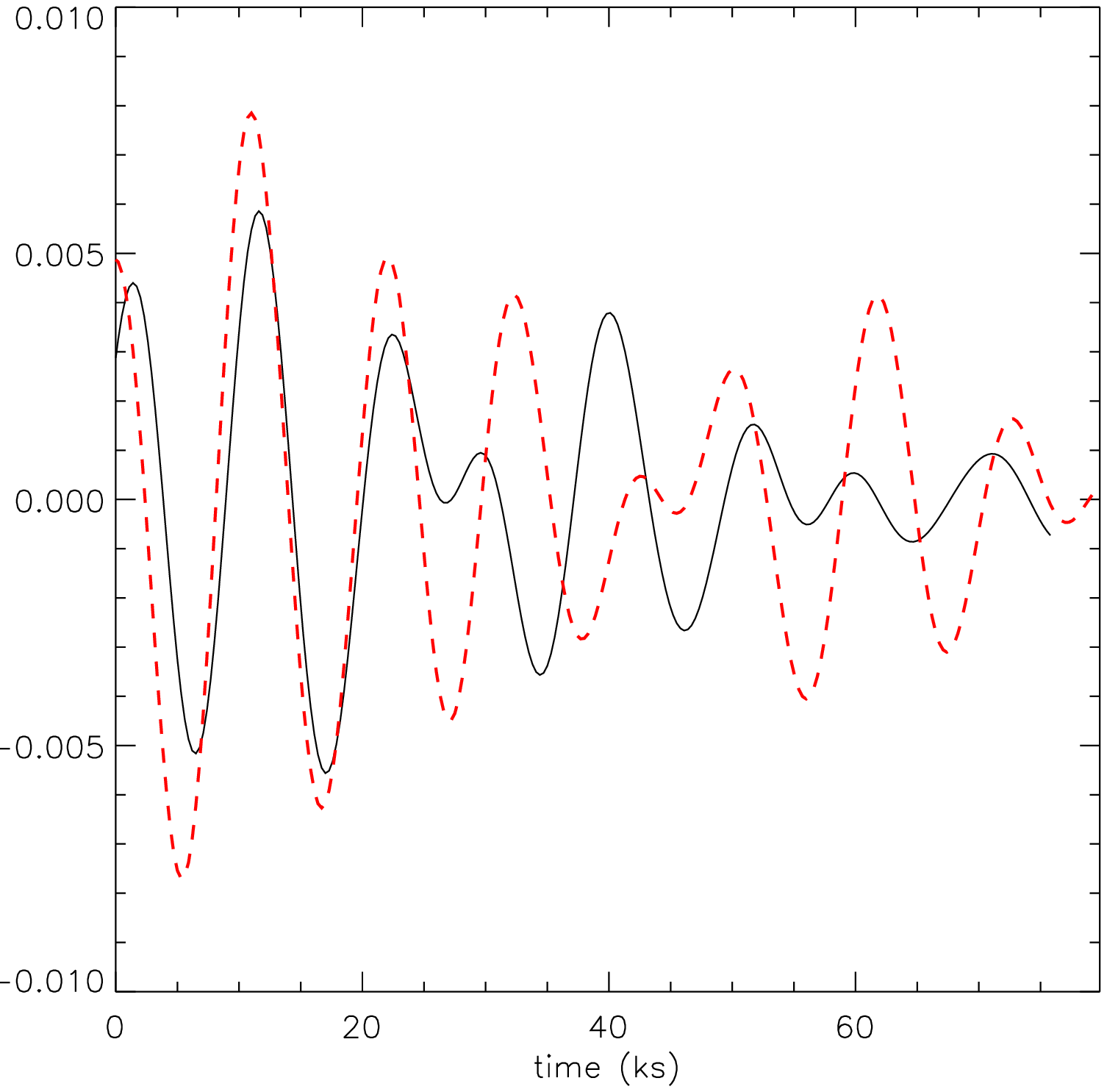}}
 \subfigure[]
   {\includegraphics[width=6cm]{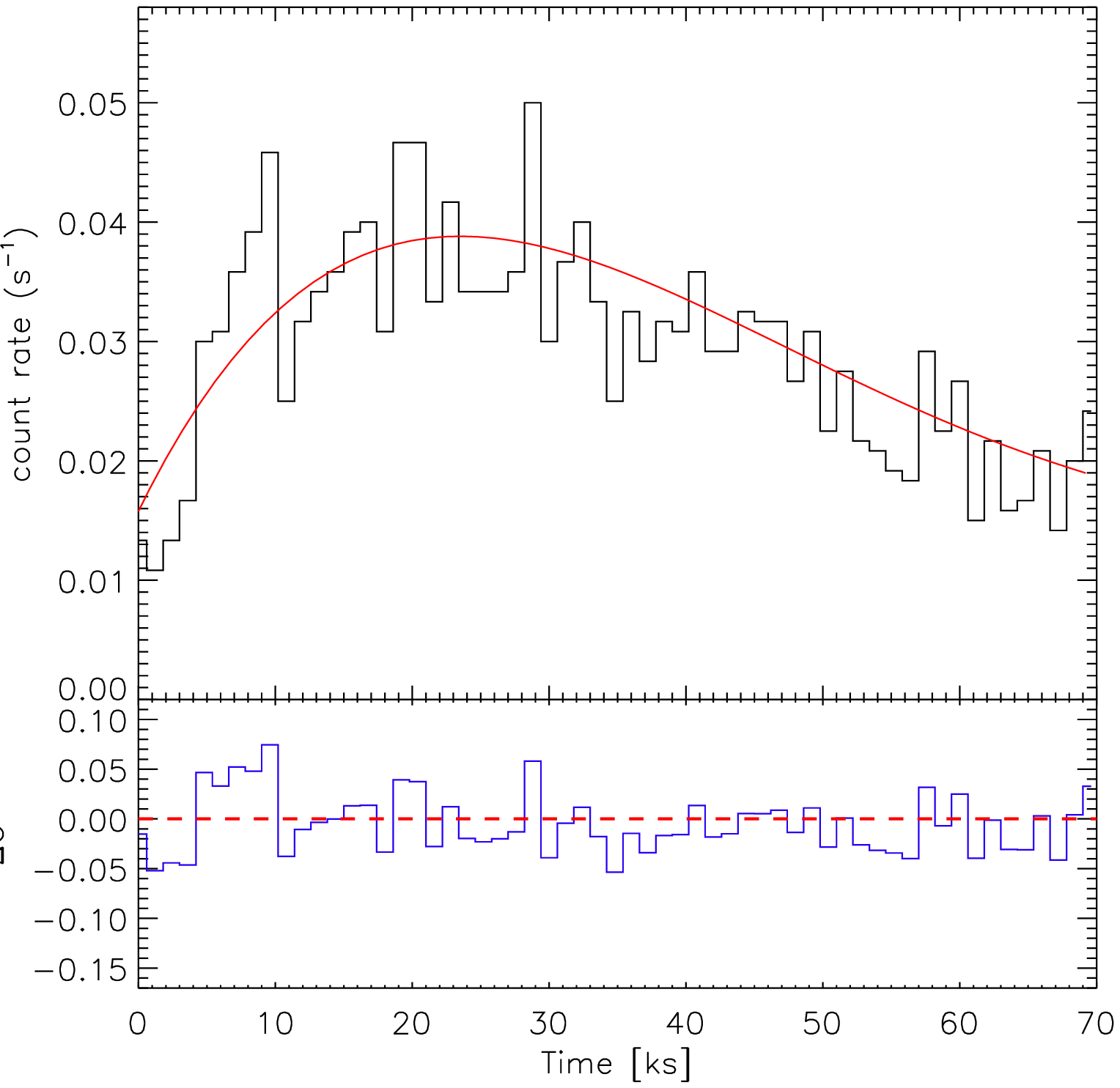}}   
 \subfigure[]
   {\includegraphics[width=6cm]{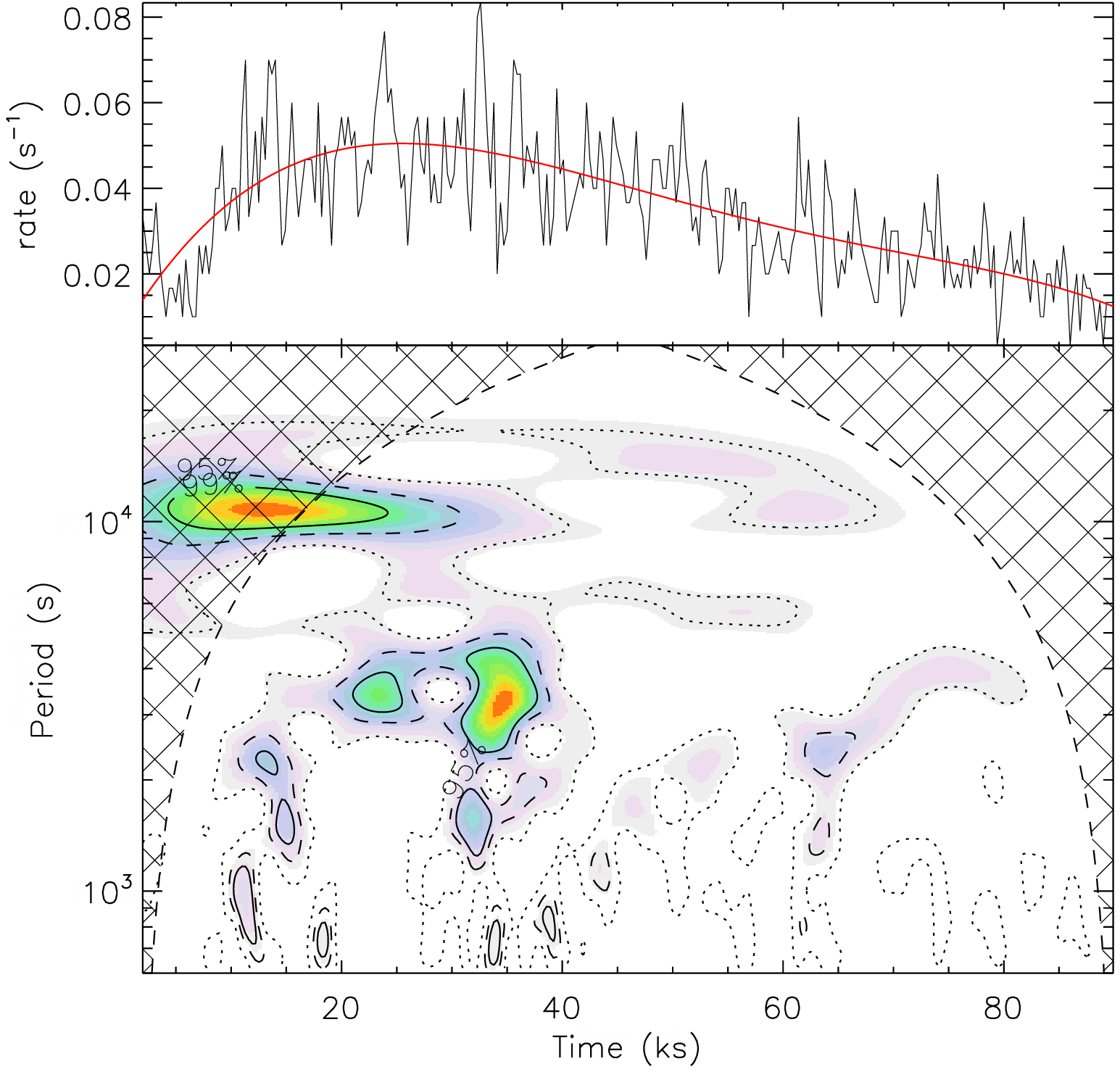}}
\subfigure[]
   {\includegraphics[width=6cm]{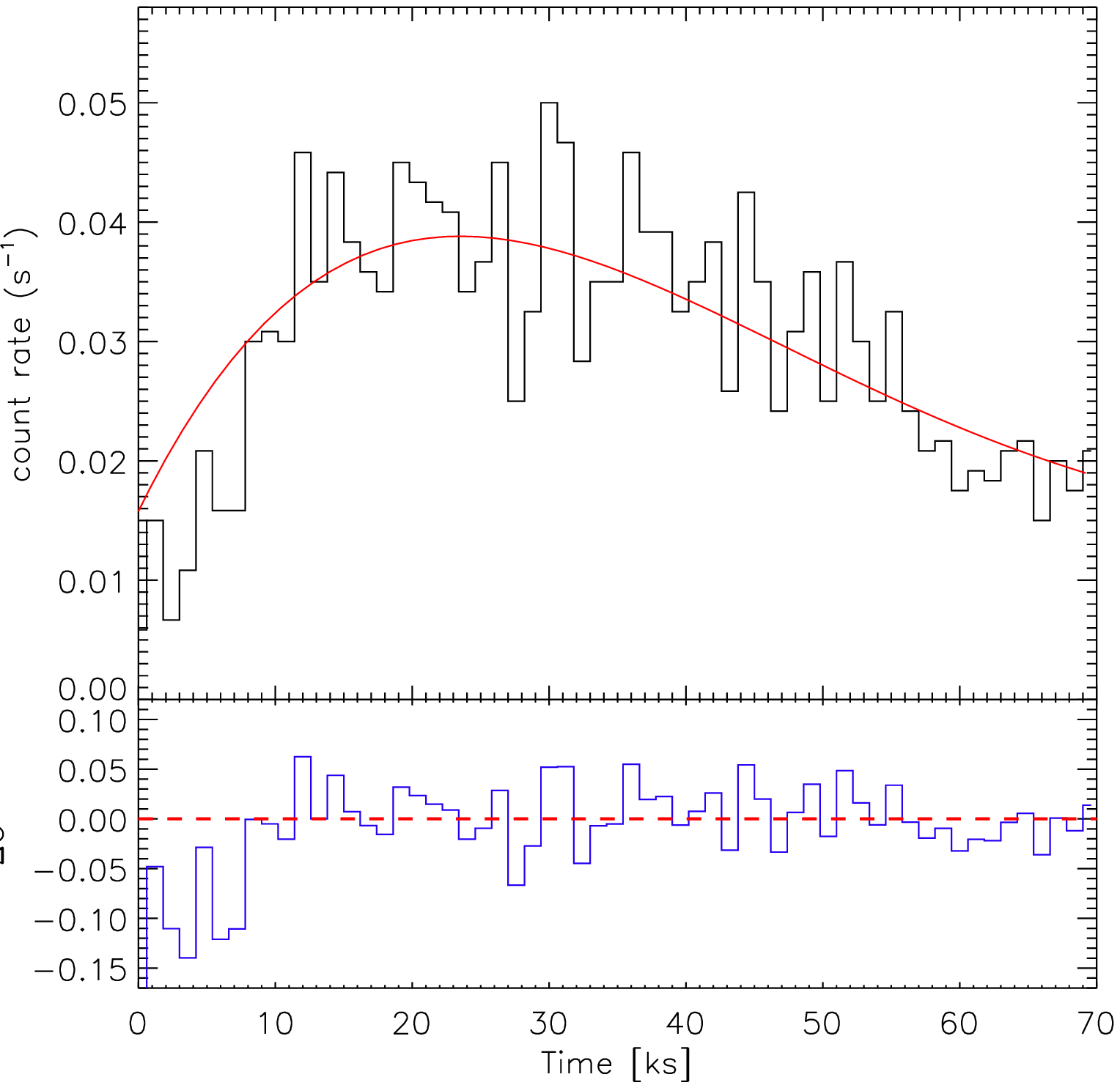}}
\subfigure[]
   {\includegraphics[width=6cm]{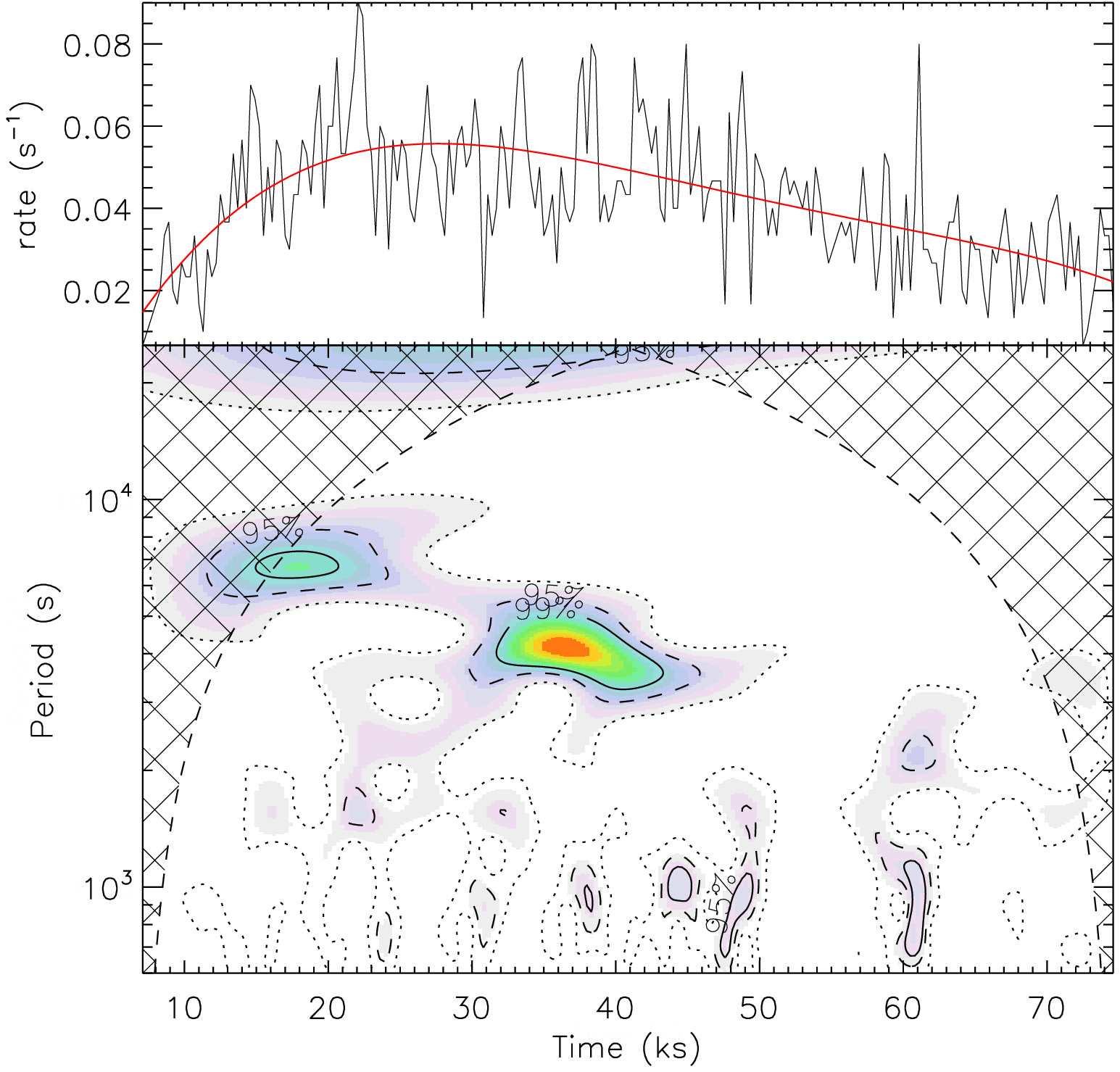}}
\caption{\footnotesize \newb{Wavelet analysis of simulated light curves: 
{(a)Observed V772 Ori flare light curve ({\it histogram}, one of those in Fig.~\ref{fig:lc}a), fitted polynomial baseline ({\it red line}), and related $\Delta C$ statistics. The temporal bin is 1.2~ks. (b) Wave reconstruction from the data (a, {\it black}) and model (c, {\it red}). (c,e) Same as (a) synthesized from the HD simulation with and without pulsations, respectively, and including Poisson noise. (d,f) Wavelet power spectrum for the light curves synthesized from the model with (c)  and without (e) pulsations, respectively.}}}
\label{fig:pois}
\end{figure}

\begin{figure}               
 \centering
   {\includegraphics[width=14cm]{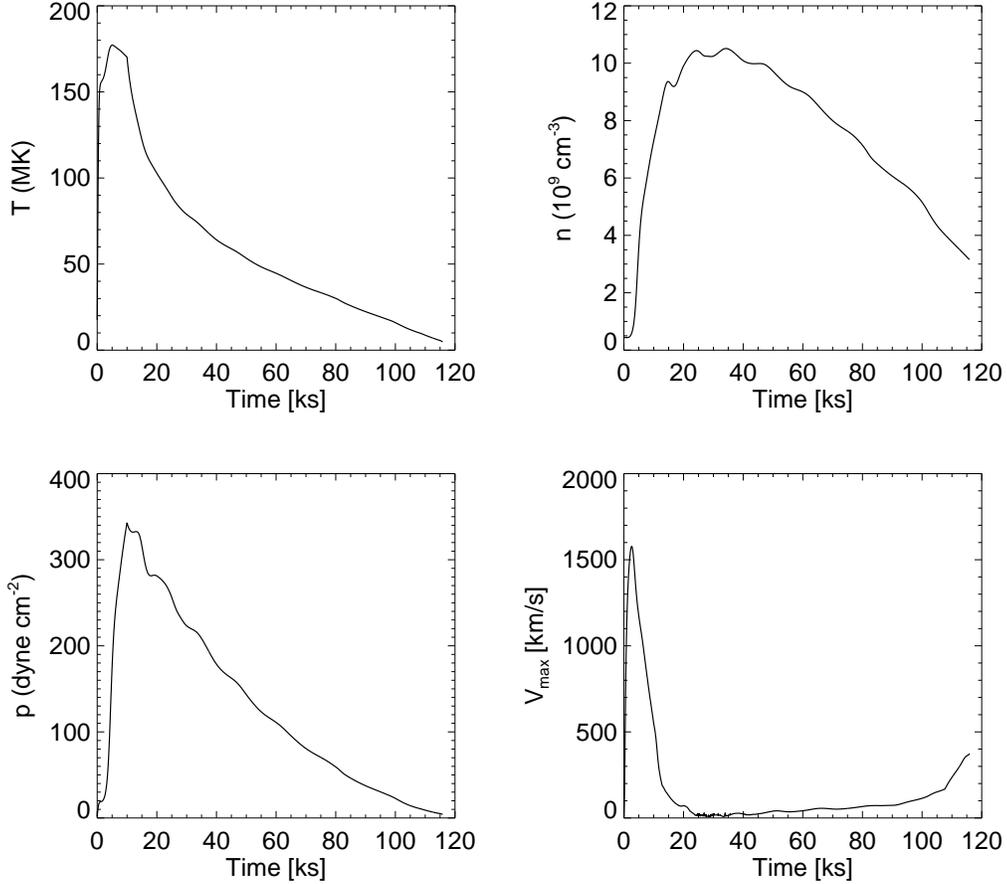}}
\caption{\footnotesize As Figure~\ref{fig:top} for a flare simulation with a heat pulse longer than $\tau_s$ which does not show pulsations. }
\label{fig:top_long}
\end{figure}

\begin{figure}               
 \centering
   {\includegraphics[width=12cm]{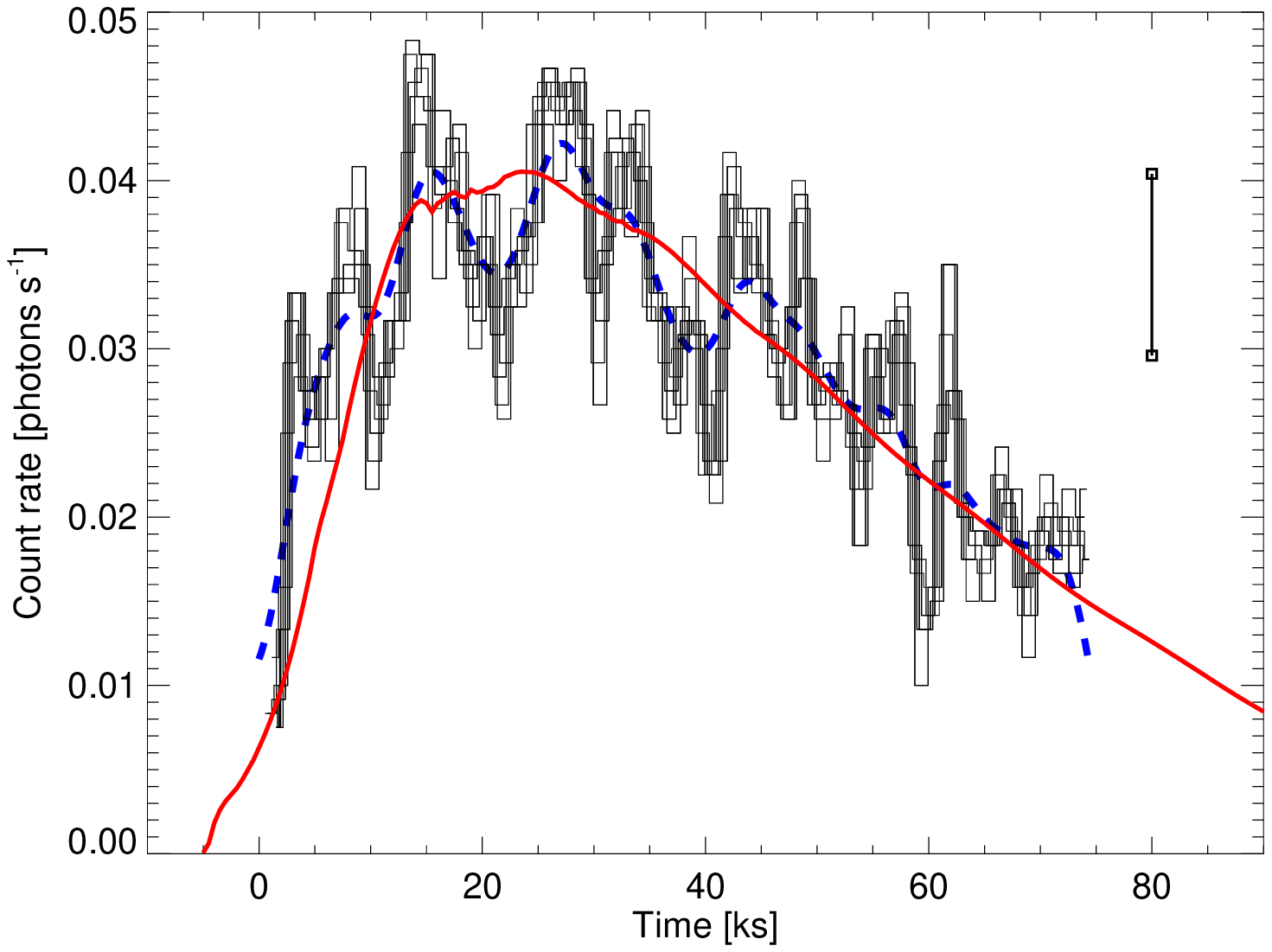}}
\caption{\footnotesize Observed light curve as in Figure~\ref{fig:lc}a (V772~Ori) compared to the one from a simulation with a heat pulse longer than~$\tau_s$ which does not show pulsations. }
\label{fig:lc_long}
\end{figure}

For completeness, we show  also results for a simulation with a heat pulse duration $\tau_H = 10000$~s ($\approx 3$~hours), i.e., longer than $\tau_s$ (Eq.\ref{eq:sct}), deposited in the same magnetic tube. In this case the heat pulse is deposited uniformly along the flux tube with a volume rate 0.05~erg~cm$^{-3}$~s$^{-1}$. Figures~\ref{fig:top_long} and~\ref{fig:lc_long} show that the overall evolution is very similar to that with the shorter pulse (Figures~\ref{fig:top} and~\ref{fig:lc}), except that there is no significant pulsation. This confirms that the pulse duration is  critical to reproduce the observed periodic pulsations  \citep{Reale2016a}. We add Poisson noise to the model light curve also for this case (Fig.\ref{fig:pois}e) and apply the wavelet analysis (Fig.\ref{fig:pois}f). The related power spectrum does not show the persistent long-period feature at $P \sim 10$ ks, as expected; \newc{the feature at $\sim 40$ ks is due to a train of tiny ripples (the amplitude is less than 1\%), not visible in Fig.\ref{fig:lc_long}, from small inaccuracies in the calculation of the X-ray emission, and is not of interest for this work.} We have also checked that the cross-correlation between the observed and the model light curves grows from 0.53 for the model without oscillations to 0.67 for the model with the oscillations (the threshold for significance is 0.25). 

Figure~\ref{fig:disk} (and Movie~1) shows how the Chandra emission predicted by the model with oscillations is distributed along the possible star-disk connecting flux tube and how it evolves in time during the modelled flare. When integrated along the tube at each time this emission corresponds to the light curve in Figure~\ref{fig:lc}. {A good \newb{agreement} is obtained by assuming that the tube is 85\% visible. Since the plasma is optically thin we expect no other effects due to system inclination. }

{For the flare on OW~Ori, we use the same model and parameters as for the one on V772~Ori, except that the length of the magnetic tube is $L_\odot = 28.6 2$, corresponding to $\sim 16$ $R_\star$. We obtain very similar results and the light curve in Figure~\ref{fig:lc}b is obtained by assuming that the tube is 70\% visible. }

\section{Discussion and conclusions}
\label{sec:discus}

\newb{The light curves in the COUP observation of V772 Ori and OW Ori  (Figure~\ref{fig:lc}) show the relatively fast rise and slower decay trend of typical flares, but the wavelet analysis detects that they are modulated by periodic pulsations. The data are compatible with a smooth wave-like modulation.} There are two key features in these pulsations: their amplitude is large, i.e. $\sim 10-20$\%, and the period is very long, i.e., $\sim 3$~hours. 
These features are detected also in other flares on Orion PMS stars \citep{Lopez-Santiago2016a}, but \newb{in these two flares they are particularly significant.} 
 
Quasi-Periodic-Pulsations (QPP) have been extensively observed in flares both on the Sun and on other stars. They have been generally connected to the propagation of waves in the flaring region \citep{Nakariakov2009a,Kumar2015a}, but ``Despite the many observational and theoretical advances in the last years, it has not been possible to determine what physical mechanism is responsible for causing the QPPs''  \citep{Van-Doorsselaere2016a}. The large amplitude makes them difficult to interpret in terms of usual MHD waves or modes. Also the periods of such modes are typically much shorter than those described here  \citep{Nakariakov2009a}.
Very recently, it has been proposed that such pulsations are caused by plasmas sloshing back and forth in closed magnetic tubes triggered by a very short and intense heat pulse \citep{Reale2016a}. This model is supported by other works \citep{Su2012a,Fang2015a} and naturally explains the large amplitudes.

%
%
%
%


Hydrodynamic simulations of the flares along a long magnetic tube at high spatial resolution show that this scenario is able to reproduce the \newb{inferred} patterns of the light curves. Figures~\ref{fig:lc} \newb{and \ref{fig:pois}a} (red solid lines) show the Chandra/ACIS light curves obtained from the simulation of a flare triggered by a heat pulse with a duration of $\sim 1$~hour \neww{inside magnetic tubes $\approx 20~R_\odot$ and $\approx 28~R_\odot$ long, respectively, which correspond to 6.7 $R_\star$ and 16 $R_\star$ for these stars, respectively (see Section~\ref{sec:data})}.

The heat pulse is spread along the tube which is symmetric with respect to the middle point, and across the tube on an area equivalent to $\sim 10$\% of the solar surface in both cases.  The pulse heats the plasma in the tube rapidly ($\sim 1$~hour) up to a temperature $\geq 200$~MK. After the heating is over the plasma rapidly cools down back below 100~MK in about 3~hours, and the cooling continues more gradually for the next several hours. The amount of plasma inside the tube has a slower evolution: dense plasma rises from the tube footpoints initially at supersonic speed ($\sim 2000$~km/s), and takes a couple of hours to fill in the tube completely. Then, the plasma is pulled back by a depression low in the tube and begins to slosh back and forth along the tube. Figure~\ref{fig:pimg} shows the sloshing for V772~Ori. After several hours the plasma starts to drain down and the emission gradually decays. 
Since the waves are hydrodynamic and the propagation is sonic, this model does not involve a direct effect of the magnetic field other than that of a waveguide. \new{Because of energy losses, the waves progressively reduce their amplitude and their period becomes longer.}

As shown in Figures~\ref{fig:lc} \newb{and \ref{fig:pois}}, this model is able to reproduce both the amplitude and the period of the pulsations of the observed flare light curves. In addition, the model \newb{is able to explain} at the same time the modulation and the envelope of the light curves (rise and decay) which are ruled by different physical processes, i.e., the sloshing and the heating and cooling, respectively. Two independent lines of reasoning therefore converge to a coherent scenario of a flare in a single and very long magnetic tube, long enough to connect the star to an accretion disk \citep{Hartmann2016a}. {As customary for time-dependent models, the model does not pretend a perfect match with the observation, because fine-tuning is prohibitive \newb{and the data do not allow to resolve enough the fine details of the light curve}. Many possible effects simultaneously present, including perturbations, geometric details, and data noise concur to determine the differences. The model reproduces well the pattern, amplitude and period of the pulsations \newb{as far as it is allowed by the observation}.}

Although QPPs are customarily observed in solar flares, the interpretation is not unique. They have already been attributed to episodic outflows \citep{Su2012a}, \new{and the model with pulse-driven sloshing is able to explain the pulsations in detail including the large amplitudes \citep{Reale2016a}. }
Similar pulsations have been found also in other numerical loop modeling \citep{Nakariakov2004a,Tsiklauri2004a,Bradshaw2013a,Fang2015a}. \new{In a more general MHD framework, the sloshing fronts can be viewed as low-order modes of slow magnetosonic waves in a low~$\beta$ plasma. The confinement of the plasma with a pressure of a few hundreds dyne cm$^{-2}$ (Fig.~\ref{fig:top}) requires a magnetic field of $\sim 100$ G, which might not be unrealistic even at such large distances from the stellar surface, for stars with average fields of a few kG at the surface \citep{Yang2011a}. Taken for granted that any kind of fast wave would imply even longer magnetic tubes, it remains the alternative possibility that the pulsations might mark slow magnetosonic waves in a high~$\beta$ environment. This cannot be completely excluded, although very unlikely because it would imply a strong coherent deformation of the magnetic field, like a fattening or shrinking of the whole tube altogether, not easy to imagine on such huge spatial scales, and also much more energy-demanding.}

\new{Our analysis shows that large and slow coherent pulsations in the day-long light curves are explained very well if flaring plasma sloshes back and forth in a single and very long magnetic tube. }
\neww{A multiloop flare \citep{Warren2006b,Rubio-da-Costa2016a} is expected to be chaotic and cannot reproduce such simple oscillation patterns, unless coherence is forced by unknown mechanisms. Nothing like this has ever been observed on the Sun, where multiloop flares are frequently observed. Our model naturally reproduces, and quantitatively, both the oscillation pattern and the envelope light curves. }

The diagnosed lengths are typical of magnetic tubes connecting the star and the inner disk \citep{Hartmann2016a}, as proposed in Figure~\ref{fig:disk} (see Movie~1). 
The framework is a young stellar object surrounded by an accretion disk according to an analytical solution \citep{Romanova2002a}. For simplicity, the magnetic field is a dipole centered on the star and perpendicular to the plane of symmetry of the disk.
The flare X-ray emission \neww{is taken from the simulation for V772 Ori and} has been mapped along a magnetic tube that links the star to the disk and $\approx 20 R_\odot$ long. The length of the magnetic tube might easily connect the star to a more distant location on the disk. This distance is well beyond the co-rotation radius, which is $\sim 1.6 R_*$ for this star. \neww{Therefore, the far footpoint is practically at rest with respect to the footpoint on the star. Since the flare duration is a significant fraction of the stellar rotation period ($\sim 180^o$), one may wonder if the magnetic tube might be significantly stretched during the flare and invalidate the model. However, one footpoint of the tube would be simply dragged along the stellar surface, and therefore the stretching would be of 2 stellar radii at most, only if the footpoint is located at the star equator. If instead it is more likely close to one of the poles, the length would be almost unchanged and the net effect would only be a twisting of the magnetic channel, which is expected and might even trigger the flare. The model is therefore fully consistent with the geometry and dynamics of the star-disk system.}

\new{In the end, this work shows \newb{strong evidence} that magnetic tubes as long as to connect the star to the circumstellar disk exist and produce flares, and even constrains the heat pulse released in the tube.} In such a scenario, flares might play a direct role also in perturbing the circumstellar disks.  In addition to obvious strong local ionization and dust destruction due to extremely hot plasma brought close to the disk, the flares produce thermal and ram pressure fronts on the order of some hundreds dyne~cm$^{-2}$ (see Section~\ref{sec:model}), comparable to the internal pressure of T Tauri disks (at an age of $10^5-10^6$ yrs) \citep{Ruden1991a}. By hitting the disk, these fronts would produce significant dynamic perturbations, e.g., warping waves, which might propagate along the disk far from the flare site and influence other processes, including those leading to planet formation. Furthermore, they might themselves trigger accretion episodes, as predicted by detailed MHD modeling \citep{Orlando2011a}.

\new{More sensitive observations of flares on PMS stars with forthcoming X-ray missions, such as Athena, will be crucial to better detect this kind of evolution and to shed more light into this scenario.}

%

\begin{figure}              
 \centering
   {\includegraphics[width=7cm]{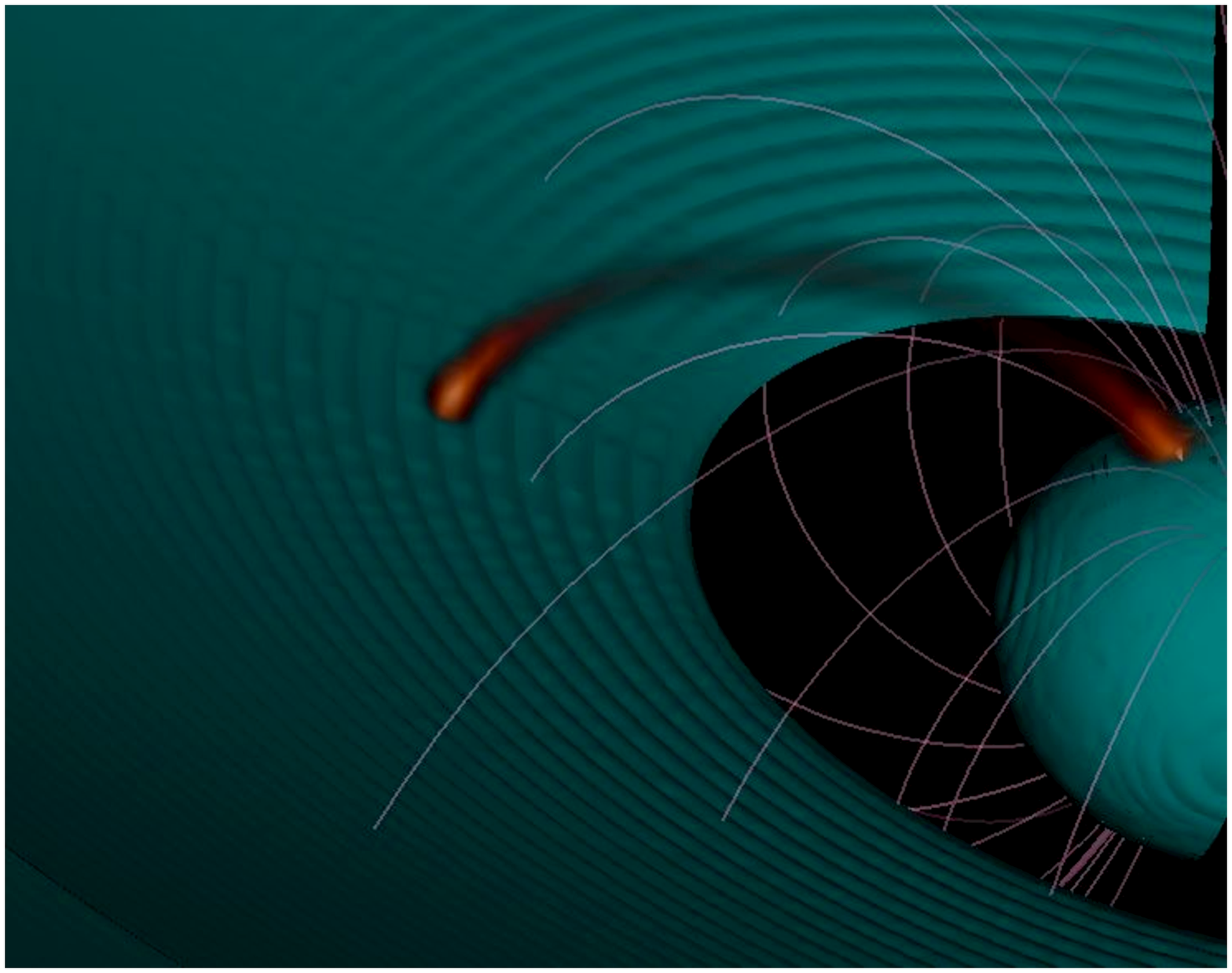}}
   {\includegraphics[width=7cm]{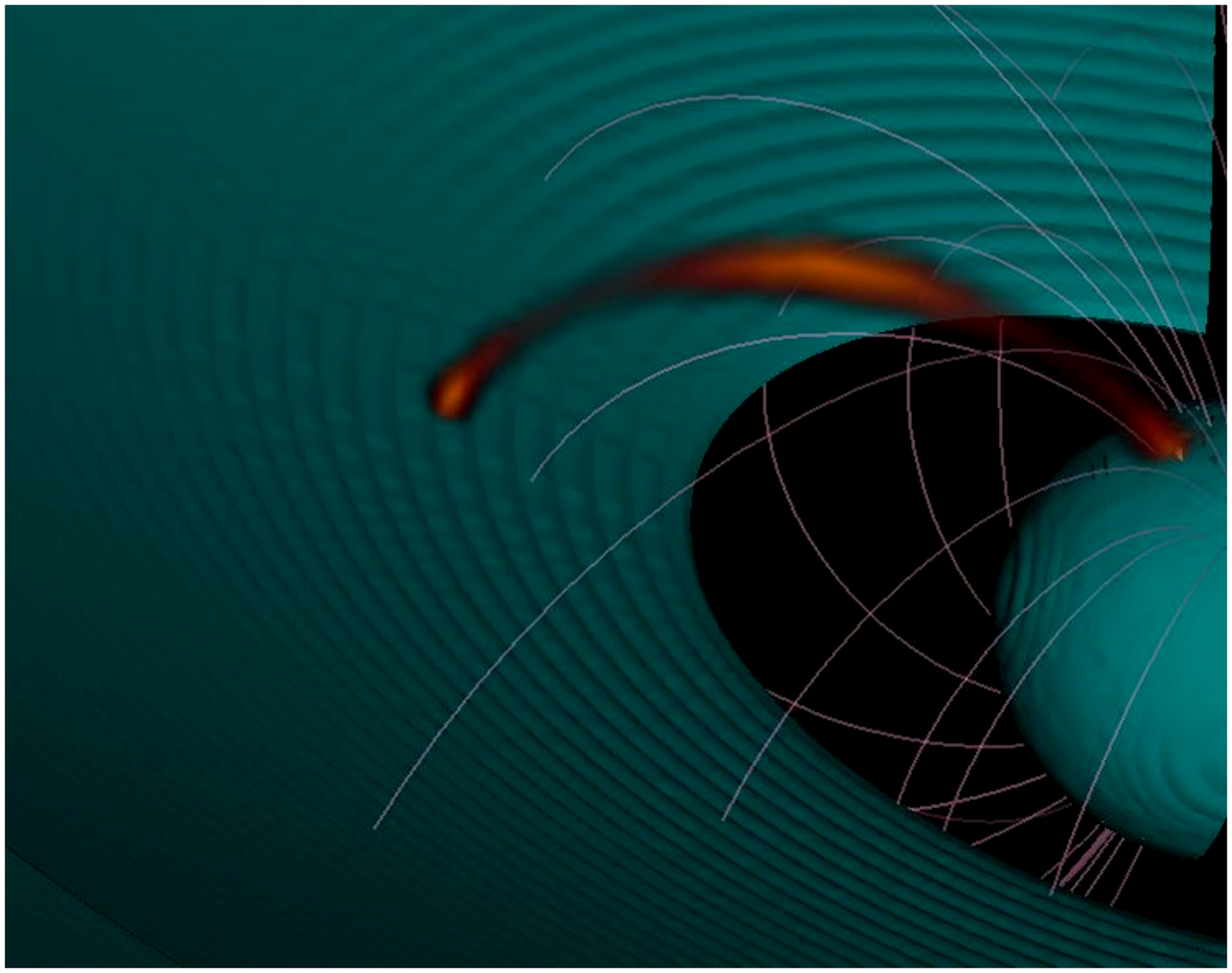}}
   {\includegraphics[width=7cm]{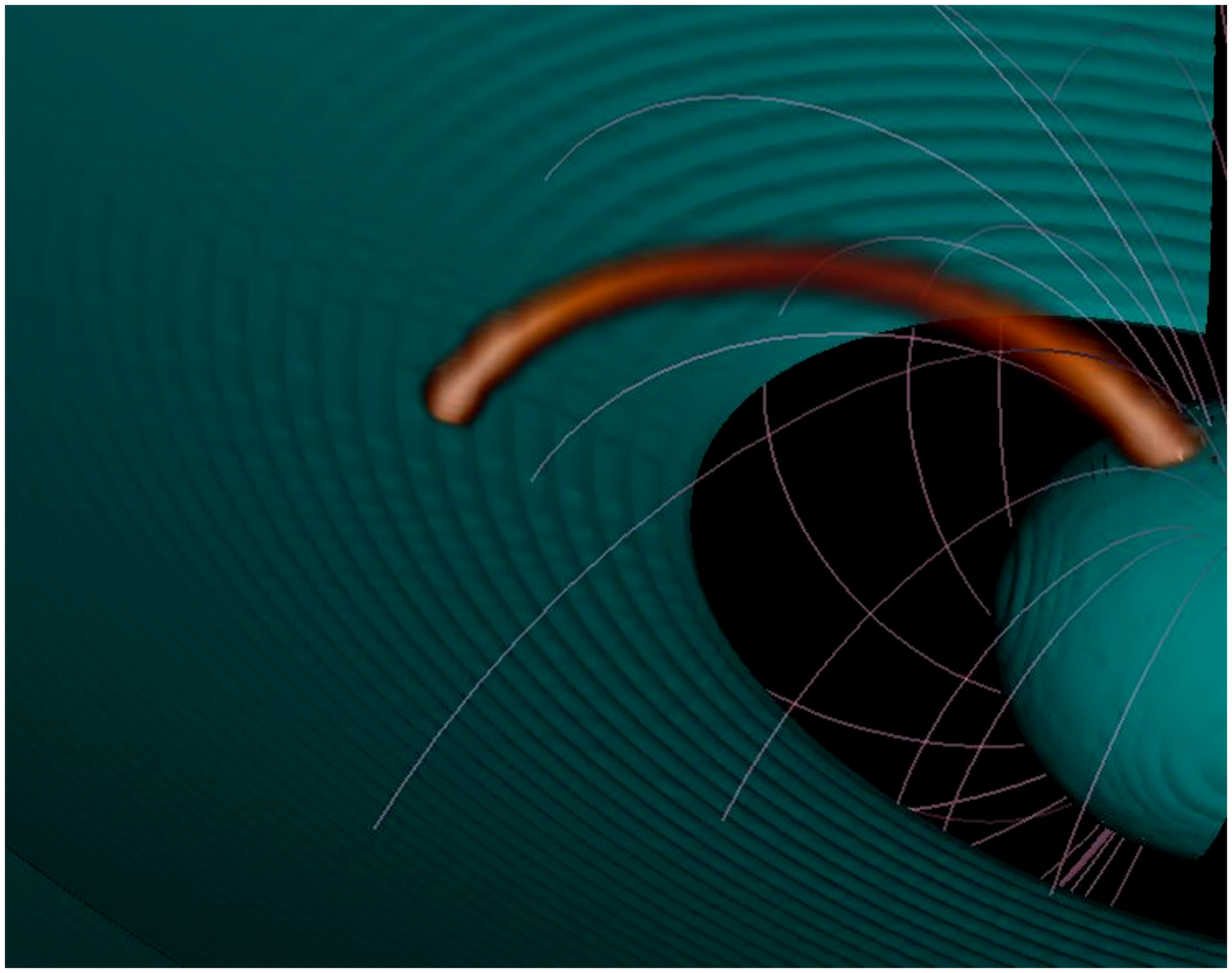}}
   {\includegraphics[width=7cm]{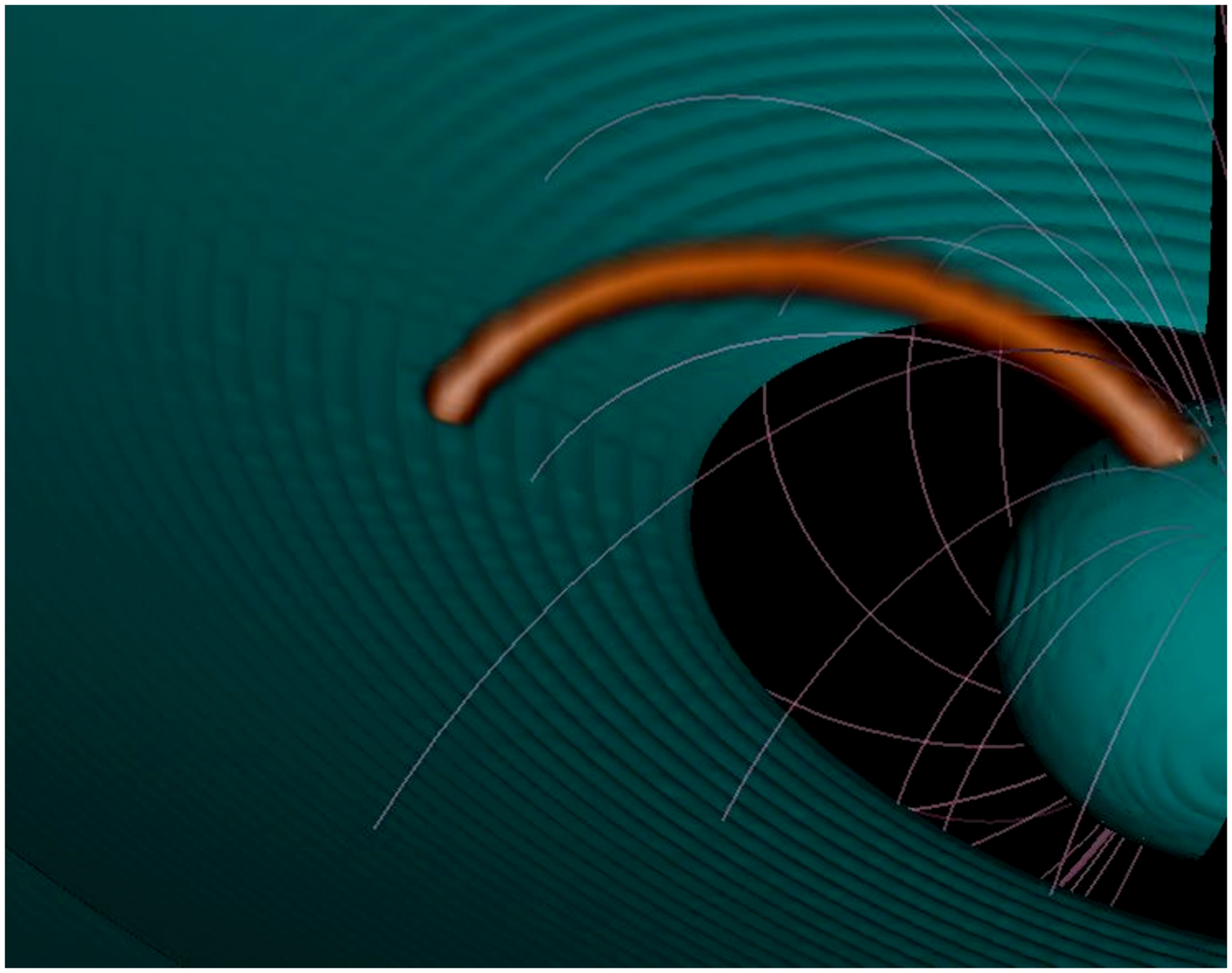}}
\caption{\footnotesize Possible scenario of the flaring magnetic tube in the V772 Ori flare. The framework is a young stellar object surrounded by an accretion disk \citep[green,][]{Romanova2002a} with a bundle of magnetic field lines (white lines). The flare X-ray emission from the simulation is mapped along a tube around a magnetic field line with a constant cross section (volumetric rendering, linear red scale) that links the star to the disk and $\approx 20 R_\odot$ long. Each frame shows the X-ray emission detectable with Chandra/ACIS, according to the hydrodynamic simulation shown in Fig.~\ref{fig:lc}. \new{The frames are taken at time 4000 s, 7000 s, 11000 s, and 15000 s since the beginning of the simulation and show the brightness fronts moving back and forth along the tube (see Movie~1 for an animated version of this figure).}}
\label{fig:disk}
\end{figure}

\acknowledgments{F.R., E.F., A.P., S.S. acknowledge support from  Italian Ministero dell'Istruzione, dell'Universit\`a e della Ricerca. J.L.-S. acknowledges the Office of Naval Research Global (award no. N62909-15-1-2011) for support. The research leading to these results has received funding from the European Union’s Horizon 2020 Programme under the AHEAD project (grant agreement n. 654215). The authors thank C. Argiroffi and F. Damiani for help.}
 

\software{XSPEC \citep{Schafer1991a}, MEKAL \citep{Phillips1999a}}

\bibliography{refs}


\end{document}